\documentclass[prl,floatfix,twocolumn,showpacs,amsmath,amssymb]{revtex4}
\usepackage{amsmath}
\usepackage{graphicx}
\usepackage{epsfig}
\usepackage{color}

\newlength{\upit}\upit=0.1truein

\newcommand{\ltappr}{{{\lower4pt\hbox{$<$} } \atop \widetilde{ \ \ \ }}}
\newlength{\bxwidth}\bxwidth=1.5 truein

\newlength{\figwidth}
\newlength{\shift}
\newcommand \bea {\begin{eqnarray} }
\newcommand \eea {\end{eqnarray}}

\newcommand{\beg}{\begin{equation}}
\newcommand{\en}{\end{equation}}

\begin{document}

\newcommand{\dg}{^{\dagger }}
\newcommand{\bfr}{\mathbf r}
\newcommand{\veps}{\varepsilon}

\newcommand{\vecr}{{\bf r}}
\newcommand{\vecz}{{\bf z}}
\newcommand{\vecq}{{\bf q}}
\newcommand{\vecp}{{\bf p}}
\newcommand{\vecs}{{\bf s }}
\newcommand{\vecl}{{\bf l}}
\newcommand{\veck}{{\bf k}}
\newcommand{\vecM}{{\bf M}}
\newcommand{\vecF}{{\bf F}}
\newcommand{\vecA}{{\bf A}}
\newcommand{\vecB}{{\bf B}}
\newcommand{\vecJ}{{\bf J}}
\newcommand{\vecG}{{\bf G}}
\newcommand{\vecsigma}{{\bf \sigma}}
\newcommand{\vecnabla}{{\bf \nabla}}
\newcommand{\vR}{{\vec{R}}}
\renewcommand{\vr}{{\vec{r}}}
\newcommand{\vj}{{\vec{j}}}
\newcommand{\vF}{{\vec{F}}}
\newcommand{\vk}{{\vec{k}}}
\newcommand{\vK}{{\vec{K}}}
\newcommand{\vq}{{\vec{q}}}
\newcommand{\vQ}{{\vec{Q}}}
\newcommand{\vPhi}{{\vec{\Phi}}}
\newcommand{\vS}{{\vec{S}}}
\newcommand{\vnabla}{{\vec{\nabla}}}
\newcommand{\vl}{{\vec{l}}}
\newcommand{\vJ}{{\vec{J}}}
\newcommand{\cG}{{\cal G}}
\newcommand{\cF}{{\cal F}}
\newcommand{\cT}{{\cal T}}
\newcommand{\cH}{{\cal H}}
\newcommand{\cJ}{{\cal J}}
\newcommand{\cD}{{\cal D}}
\newcommand{\cL}{{\cal L}}
\newcommand{\Tr}{\mathrm{Tr}}
\renewcommand{\a}{\alpha}
\renewcommand{\b}{\beta}
\newcommand{\g}{\gamma}
\renewcommand{\d}{\delta}
\renewcommand{\Im}{\textrm{Im}}
\renewcommand{\Re}{\textrm{Re}}
\newcommand{\cA}{{\cal A}}

\title{Neutron magnetic form factor in strongly correlated materials}

\date{\today}
\author{Maria Elisabetta Pezzoli$^{1}$, Kristjan Haule $^{1}$, 
 and Gabriel Kotliar$^{1}$}
\affiliation{
$^{1}$Serin Physics Laboratory, Rutgers University,Piscataway, NJ 08854, USA. \\}

\begin{abstract}
We introduce a formalism to compute the neutron magnetic form factor
$\vecF_{{M}}(\vecq) $ within a first-principles Density Functional
Theory (DFT) + Dynamical Mean Field Theory (DMFT). The approach treats
spin and orbital interactions on the same footing and reduces to
earlier methods in the fully localized or the fully itinerant limit.
We test the method on various actinides of current interest
NpCoGa$_5$, PuSb and PuCoGa$_5$, and we show that  PuCoGa$_5$
is in mixed valent state, which naturally explains the measured
magnetic form factor.
\end{abstract}


\maketitle

Compounds including elements from the actinide series provide a beautiful illustration
of the challenges posed by correlated materials. The  $5f$ electrons in these systems display
simultaneously itinerant (i.e.  band-like)  and localized (atomic-like) properties.
Describing the impact of this wave-particle duality  on different physical observables, measured
using different spectroscopic probes, is an  outstanding theoretical challenge.

Neutron scattering\cite{Lovesey}  is a time-honored probe to
investigate the dynamics of the magnetic degrees of freedom. It probes
the dynamic susceptibility, describing the spatial and temporal
distribution of magnetic  fluctuations. In the itinerant limit,
it can be modeled in terms of a  particle hole continuum of
quasiparticles, while in the localized limit it can be describe
in terms of propagating spin waves. It is generally accepted that
in many materials neither a fully itinerant nor a fully
localized picture is adequate and some combination of both is
required to model the dynamics of the spin fluctuations  as in the
duality model of Ref.~\onlinecite{Miyake199120}. \\
The intensity in the magnetic Bragg peaks 
can be used to obtain a
real picture of the magnetization inside the unit cell. This can
be done even for materials that do not exhibit magnetic long
range order, by applying an external magnetic field.
 Classical techniques can handle a fully itinerant or a fully localized
picture \cite{raree}.
However these approaches are not sufficient for many compounds of
considerable scientific interest. It has been known for a while 
that intermediate valence rare-earth semiconductors show puzzling 
magnetic properties that can be explained only by a theory 
which explicitly considers the spatial extend  of the magnetic 
excitations \cite{Kikoin}.
Similarly only magnetic orbitals of strong covalent nature can correctly
account for the neutron intensity in the cuprates  \cite{Savici}.
A theory able to describe the magnetic form factor for partly itinerant
systems from first principles is needed.

Important recent experiments of Hiess \emph{et. al.} determined the
magnetic field induced form factor of PuCoGa$_5$, a  material
which superconducts at  the remarkably high  transition
temperature $T_c \simeq 18.5 \; K$, a record  in the  heavy-fermion
family \cite{Hiess}. The degree of itinerancy of the f electrons is the subject
of active debate and has important consequences for the mechanism
of superconductivity. Neither the localized nor the itinerant
model of the neutron form factors fits the data well, providing
strong motivation for our theoretical developments.

In this letter we develop a method to compute the form factor for
magnetic neutron scattering within DFT+DMFT~\cite{review}. We test the
method on several actinide materials.  The PuCoGa$_5$ induced magnetic
form factor is consistent with correlated mixed valent nature of the
material, where both the $5f^5$ and $5f^6$ configuration are
important. This is reminiscent of the mixed valent nature of elemental
plutonium~\cite{Shim}.

The magnetic form factor $\vecF_M(\vecq) $ is defined by
\beg
\vecF_M(\vecq)   = - \frac{1}{2 \mu_B}  \Big \langle  \vecM_T(\vecq) \Big \rangle \,,           
\en
where $\vecM_T(\vecq) = \vecM^{\, \textrm{spin}}_T(\vecq) + \vecM^{\, \textrm{orb}}_T(\vecq)$ 
is the Fourier transform of the transverse component of the magnetization density
$  \hat{q} \times \!  \big ( \vecM(\vecr) \times \hat{q} \big) $, $\mu_B$ is the Bohr magneton
and $\vecq $  is the scattering wave vector at the Bragg peak.
To avoid ambiguity in definition of magnetization~\cite{Hirst}, we
express the form factor in terms of the Fourier transform of the
current density $\vecJ(\vecq)=\int \textrm{d} \vecr \; e^{-i \vecq
  \cdot \vecr} \vecJ(\vecr)$.  The current and the transverse
magnetization are related by $ \vecM_T(\vecq)= \frac{i}{c} \vecq
\times {\vecJ(\vecq)}/{q^2}$. The current has two contributions, the
spin part $\vecJ_{\textrm{spin}}(\vecr)$ and the orbital part
$\vecJ_{\textrm{orb}}(\vecr)$.
Expressing the definition of $\vecJ_{\textrm{orb}}(\vecr) $ and
$\vecJ_{\textrm{spin}}(\vecr) $ in terms of field operators
$\Psi_s(\vecr)$, we find for the form factor the following expression
\beg
\begin{split}
\label{eq:2quant}
\vecF_M(\vecq) &= \frac{1}{q^2}
\sum_{s s'} \int
\!   \!\textrm{d} \vecr \; e^{-i \vecq \cdot \vecr}\times
\\
&
\Psi_s^{\dagger}(\vecr)\; \vecq \times 
\left[ \frac{1}{2} \vec{\sigma}_{s s'} \times \vecq + \delta_{ss'} \vec{\nabla} \right] \Psi_{s'}(\vecr) \,,
\end{split}
\en where $\vec{\sigma}$ is the vector of the Pauli matrices and
$s,s'$ are the spin indexes.
For a more detailed derivation see the on-line supplementary
material \cite{epaps}. 
  It is useful to notice that the limit
$\lim_{q \rightarrow 0} \vecF_M(\vecq) $ can be well defined,
but it is subtle~\cite{Thonhauser}.  However the form factor
$\vecF_M(\vecq) $ is measured only at finite $\vecq$ values and hence
it is free from ambiguities.

$\vecF_M(\vecq) $  is measured in polarized-neutron diffraction
experiments directly through the  \emph{flipping ratio} technique.
In this method an external magnetic field $\vecB$ is applied to the
sample, and the ratio $R =\left( d \sigma /d \Omega\right)_+ / \left( d \sigma /d \Omega\right)_- $ between the cross section
for neutrons polarized parallel and anti-parallel to $\vecB$ is measured.
In a centrosymmetric crystal structure with collinear magnetic moments 
and $\vecq \perp \vecB $  the flipping ratio $R$ satisfies 
$ \left ( \frac{\sqrt{R}-1}{\sqrt{R}+1} \right) =  {\gamma r_0 F_M(\vecq)}/{b} $
where $b$ is the known nuclear scattering amplitude, $F_M(\vecq)$ the component of the magnetic structure
factor parallel to $\vecB$, $\gamma=1.9132$ and $r_0=\hbar e^2/ m c^2$
is the classical electron radius.
More general formulas which relate the form factor to the flipping ratio for
other crystal structures and experimental setups are given in
Ref.~\onlinecite{Balcar}.
For localized electrons the form factor is commonly fitted to the 
following radial dependence 
$F_M(q) = -\frac{\mu}{2\mu_B} \big(\langle j_0(q) \rangle + C_2 \langle
j_2(q) \rangle \big)$, where $\langle j_{k} (q) \rangle$ stands for
the spatial average over the atomic wave function of the magnetic atom
(which is usually solved in the isolation).
This should be understood in the so called \emph{dipole
  approximation}.
The exponent $e^{-i\vec{q} \,\vec{r}}$ is expanded around
the center of the atom as
$e^{-i\vec{q} \,\vec{r}}\approx j_0(q r) -i
(\vec{q}\cdot\vec{r}) (j_0(q r)+j_2(q r))$, where $j_k(qr)$ are
spherical Bessel functions of order $k$.  Within this approximation,
the form factor is greatly simplified and in the common experimental set
up ($\vecq \perp \vecB$, $\vecB = B\hat{\vecz}$), it
reduces to \beg
\label{eq:6}
F_M(q) = 
\Big \langle    s_z   j_0(qr)  +  \frac{1}{2} l_z   \left \{ j_0(qr) + j_2(qr) \right \}    \Big \rangle \,.
\en
Here $r$ is the distance from the magnetic atom, and
$\langle\cdots\rangle$ stands for the spatial and temporal average.
The first and the second term in Eq.~(\ref{eq:6}) come from the spin and
the orbital contribution, respectively. 
 The comparison of the above expansion with Eq.~(\ref{eq:6}) shows that $\mu=-\mu_B\langle 2
s_z+l_z\rangle$ and $\mu C_2 =-\mu_B\langle l_z\rangle$, hence $C_2 =
\mu_L/ (\mu_L + \mu_S )$.  Clearly the ratio $C_2$, which is given by
the shape of the form factor, uniquely determines the size of the
orbital and spin component within the dipole approximation.  Even so,
caution is necessary in interpreting experiments with
Eq.~(\ref{eq:6}), since a priori the magnitude of higher order terms
beyond the dipole approximation is not known ~\cite{Rotter,Ayuel}.

To compute the form factor within DFT+DMFT, we apply a small magnetic
field $ \vecB =B \hat{{\bf z}}$ to induce a finite magnetic moment.  We solve the
DMFT problem in the presence of magnetic field, and evaluate the
correlation function Eq.~(\ref{eq:2quant}). When expressed in the
Kohn-Sham basis, Eq.~(\ref{eq:2quant}) takes the form
\begin{equation}
\begin{split}
\label{eq:ks}
\vecF_M(\vecq)& = \frac{1}{q^2}\sum_{\veck,ij,ss'} n^{\textrm{DMFT}}_{\veck,ij} \\
\times \int_{\textrm{unit cell}} \! \!  \! \! \! \! \! \! \! \!  \textrm{d}\vecr &
e^{-i\vecq \cdot \vecr}
\psi^*_{\veck i}(\vecr,s)\;\vecq\times\left[\frac{1}{2}\vec{\sigma}_{ss'}\times\vecq + \delta_{ss'}\vec{\nabla}\right]
\psi_{\veck j}(\vecr,s'),
\end{split}
\end{equation}
where $\psi_{\textbf{k}i}(\textbf{r},s)$ are the Kohn-Sham orbitals,
$i$ runs over the Kohn-Sham bands, and $\veck$  over the first
Brillouin zone.
The ``DMFT density matrix'' $n^{\textrm{DMFT}}_{\veck,ij} $ is
expressed in terms of the
DMFT Green function $G_{ij}(\veck,\omega) $ in the solid
$ n_{ij \veck}^{\textrm{DMFT}} = \frac{1}{2\pi i} \int \textrm{d}
\omega
\; \left( G^*_{ij}(\veck,\omega)-G_{ji}(\veck,\omega)\right) f(\omega)  \,,
 $
where $f(\omega)$ is the Fermi function. The form factor is thus
expressed in terms of the one particle correlation function, which is
easily accessible within DMFT. Moreover, the spatial integral is local
and runs over one unit cell, which makes local DMFT approximation
particularly suitable for this problem.
We implemented Eq.~(\ref{eq:ks}) within the recent realization of
DFT+DMFT~\cite{HauleCk} based on Linear Augmented Plane Wave (LAPW)
basis set as implemented in the full potential electronic structure
code Wien2k~\cite{Blaha}. The explicit formulas for the form factor
evaluation within this basis set, as well as detail derivation of
Eq.~(\ref{eq:ks}) are given in the on line material ~\cite{epaps}. To solve the
impurity problem in the presence of magnetic field, we used the
Non-Crossing Approximation~\cite{HauleCk}.
Our calculations show  small anisotropic corrections to dipole
approximation for the materials studied here, suggesting that the
dipole approximation is a good approximation for these compounds.
For comparison, we also compute the form factor within Local Spin
Density Approximation (LSDA)
as first discussed in Ref.~\onlinecite{Freeman}.
In practice we evaluate the mean value of Eq.~(\ref{eq:6}) 
inside the atomic sphere following the lines of  Ref.~\onlinecite{Brooks}.
We  perform the LSDA calculation in the
presence of external magnetic field, as implemented in Wien2K
\cite{ORB}. 

\begin{figure}
\includegraphics[width=0.50\textwidth]{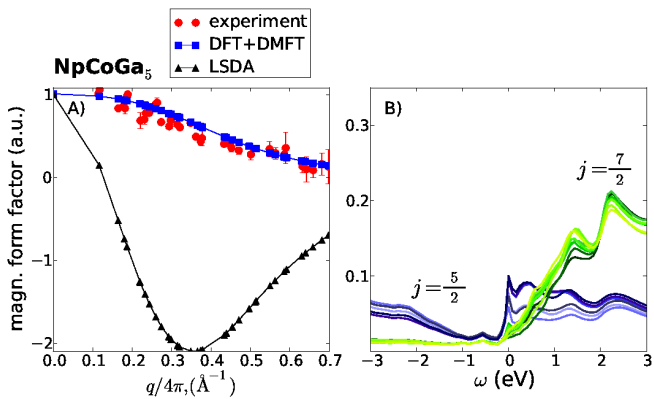}
\caption{\label{fig:NpCoGa5} (Color on line) A) panel: magnetic form
  factor for NpCoGa$_5$.  Red dots are experimental data reproduced
  from~\onlinecite{Hiess}. The blue curve with squares is the DFT+DMFT
  calculation, and the black curve with triangles is the LSDA
  calculation. The DFT+DMFT form factor agrees with experiment with
  value of the Pearson correlation coefficient
  $R_{\textrm{PMCC}}=0.95$. B) panel: spectral function $A_{j
    m_j}(\omega)$ for Np f-electrons. Blue curves correspond to the
  $j=5/2$ multiplet and green curves to the $j=7/2$ multiplet.  The
  experimental and theoretical DMFT temperature is $T=52$ K.}
\end{figure}

In Fig.\ \ref{fig:NpCoGa5}(A) we compare theoretical DFT+DMFT and LSDA
form factors with experiments on NpCoGa$_5$ in the paramagnetic
state~\cite{Hiess}.  Our DFT+DMFT form factor is in excellent 
agreement with experiment, while
the LSDA  dramatically fails in this material.
The LSDA form factor shows a minimum at finite wave vector $\vecq$.
Such a large minimum can be explained by $C_2 \sim -9.5 $;
this  occurs since  $\mu_L$ and  $\mu_s$ almost cancel,
but $|\mu_L| < |\mu_s|$. An underestimation of the orbital moment
is typical of LSDA.
Within DFT+DMFT the atomic degrees of freedom are treated exactly by
the exact diagonalization of the atomic $5f$-shell in the presence of
magnetic field. This ensures that Hund's rule coupling is properly
treated, leading to anti-parallel $\mu_L $ and $\mu_S$, but $|\mu_L |
> |\mu_S|$, hence $C_2 > 0$.
For NpCoGa$_5$ we determine the value of the coefficient $C_2=2.16$.
This value is consistent with localized $5f$-electrons in the
configuration $5f^4$, in agreement with M\"ossbauer spectroscopy
\cite{Metoki} and neutron diffraction experiments
\cite{Hiess,Colineau}.  At the same time NMR \cite{Sakai} and
inelastic neutron scattering \cite{Magnani} suggest that NpCoGa$_5$
shows also itinerant aspects of the $5f$-electrons.  A signature of
this moderate delocalization is also apparent in our calculated
spectral function at $T=52 \; K$ displayed in Fig. ~\ref{fig:NpCoGa5}(B). A
small quasiparticle peak is formed at the Fermi level, a signature of
electron itinerancy at low energy.
\begin{figure}
\includegraphics[width=0.5\textwidth]{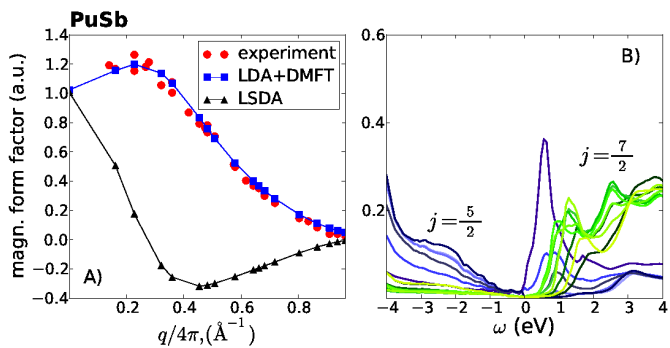}
\caption{\label{fig:PuSb} (Color on line) A) panel: magnetic form
  factor for PuSb at $T=20 \;K$. Red dots are experimental data
  ~\cite{Lander1984}, the blue curve with square is the DFT+DMFT
  calculation and the black curve with triangles the LSDA calculation.
The DFT+DMFT curve agrees with experiment with a Pearson Correlation coefficient
$R_{\textrm{PMCC}} = 0.95$.
B) panel: Spectral function for PuSb. The color legend is the same
  as in Fig.\ \ref{fig:NpCoGa5}. }
\end{figure}
Next we compute the form factor for PuSb in the ferromagnetic state.
PuSb is a  metal ~\cite{PuSbmetal},
which orders antiferromagnetically below $T_N = 85 \; K$
and becomes a ferromagnet at $T = 67 \; K$  \cite{PhysRevB.30.6660}.
Theoretically it has been showed that in PuSb valence fluctuations
are suppressed with the consequent absence of a
quasiparticle multiplet structure in the spectral function~\cite{Chuck}.
This result is consistent with neutron diffraction data:
 the form factor curve has a characteristic maximum at finite $q$, feature typical
of a pure $f^5$ configuration state for the Pu atom~\cite{Lander1984}.
The LSDA calculation underestimates the orbital moment and finds a negative $C_2$
coefficient.
Our DFT+DMFT calculation reproduces the $f$-electrons occupation
value $\langle n_f \rangle \sim 5.0$ of the previous experimental and theoretical  works~\cite{Chuck,Lander1984}
and indeed it is in good agreement with  the measured data, see (see Fig.~\ref{fig:PuSb}A).
In particular we find that  there is a large cancellation between orbital
and spin moment with $\mu_S/\mu_L = -0.74$ and $C_2 = 3.92 $.
\begin{figure}[ht]
\includegraphics[width=0.5\textwidth]{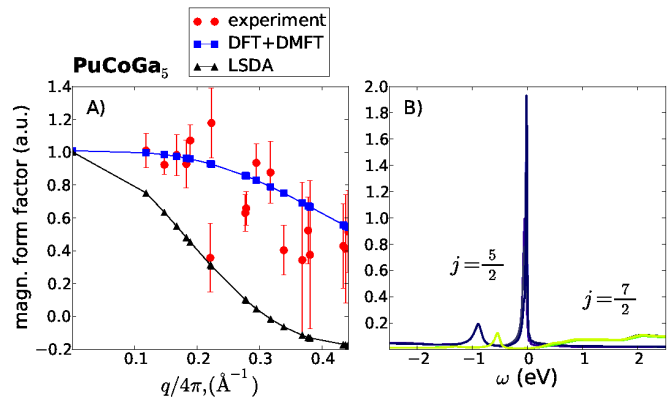}
\caption{\label{fig:PuCoGa5}(Color on line) A) panel : magnetic form
  factor for PuCoGa$_5$. The blue curve with squares corresponds to
  the full DFT+DMFT calculation, the black curve with triangles to the
  LDA calculation. Red dots are experimental data ~\cite{Hiess} The
  DFT+DMFT curve agrees with experiment with a Pearson correlation
  coefficient $R_{\textrm{PMCC}}=0.70$ .  B) panel: spectral function
  $A_{j m_j}(\omega)$ for Pu f-electrons. The color legend is the same
  as in Fig.\ \ref{fig:NpCoGa5}}
\end{figure}
We now turn to PuCoGa$_5$. Photo emission spectra show the
formation of a quasiparticle peak at the Fermi level, however there is
a large discrepancy in the peak height between different measurements
\cite{Eilordi,Joyce}.  First magnetic susceptibility measurements
suggested that $5f$-electrons behave as unquenched local moments until
they enter in the superconducting state ~\cite{Sarrao1}. In turn
neutron scattering shows a temperature independent magnetic
susceptibility, implying the absence of magnetic moments such as in
$\delta$-Pu ~\cite{Hiess,Lashley}.
Electronic structure calculations qualitatively support the picture of
delocalized $5f$-states, however they predict a Pu ion close to
magnetic order and a form factor shape not observed in experiments
\cite{Ophale1,Shick}.  Since our understanding of superconductivity in
PuCoGa$_5$ depends on the itinerant or localized nature of correlated
electrons \cite{Maxim}, further theoretical and experimental
investigations are compelling.  Within our DFT+DMFT calculation we find
that a quasiparticle peak appears at the Fermi level, see
Fig.~\ref{fig:PuCoGa5}(B). These results are consistent with a specific heat coefficient
$\gamma \sim 70 \; mJ/(K^2 mol) $, which compares well with experiments \cite{Sarrao1},
and go beyond the pioneer DFT+DMFT calculations,
solved within the T-matrix and fluctuating exchange technique \cite{Pourovskii}.
Together with a quasiparticle peak, a mixed valent state forms, where
the $5f$-electrons have a finite probability to be both in the
configuration state $f^5$ and $f^6$.
Our theoretical prediction for the $5f^6$ occupation probability is
$P_{f^6}=0.26$, corresponding to $\langle n_f \rangle \sim 5.26$ and a
coefficient $C_2= 2.35$.
We plot the corresponding form factor curve in Fig.~\ref{fig:PuCoGa5}(A)
together with the form factor obtained from the LSDA calculation.
As for the previous materials,
 LSDA underestimates the orbital moment and it
obtains a negative $C_2$ coefficient that is inconsistent with
experimental data. 
The DFT+DMFT form factor with $C_2=2.35$ well describes
the neutrons data and it accounts also for the magnetic susceptibility 
( see the supplementary material \cite{epaps}). 
 The value of $C_2=2.35$
is naturally explained by the mixed valence picture obtained
theoretically for PuCoGa$_5$. 
For a free Pu$^{3+}$ ion solved in the intermediate coupling
$C_2 =3.83$, hence $\mu_L/\mu_S=-1.83 $~\cite{raree};
As a mixture of the configuration $f^6$ is included in the
many body ground state, the ratio $\mu_L/\mu_S$  becomes more negative
and therefore $C_2$ decreases.
As pointed out in Ref.~\onlinecite{Hiess} the $F_M(q)$ shape is very
different from the one expected for a pure $5f^5$ configuration of an
isolated Pu ion, as for example is found in PuSb, see Fig.~\ref{fig:PuSb}(a). 
At the same time it is very different from the LSDA prediction.
Hence, the magnetic properties of PuCoGa$_5$ are not captured either
by a free moment picture or by an itinerant picture.
 
In conclusion in this letter we presented a new approach
to compute the neutron magnetic form factor.
The LSDA treatment fails to reproduce the 
correct form factor since the exchange energy is orbital-independent
and therefore Hund's rules are not respected.
On the contrary DFT+DMFT includes the atomic physics
needed to describe strongly correlated systems.
Application of DFT+DMFT to  PuCoGa$_5$ suggest an explanation
of the results of Ref.~\onlinecite{Hiess} in terms of a mixed valence 
picture where the ground state of Pu fluctuates between two distinct
configurations: $f^5$ and $f^6$. We indeed checked that this picture
accounts for the values of the specific heat and susceptibility as well
as for the shape of photo-emission spectra. 
We find a close similarity between the DFT+DMFT valence histogram
of PuCoGa$_5$ and $\delta-$Pu, suggesting a close analogy
of the local physics in these two materials; the magnetic form factor
of PuCoGa$_5$ would then be very similar to that of $\delta-$Pu,
for which experiments are notoriously difficult.
Finally, mixed valence  is an attractive mechanism for 
pairing in heavy fermions \cite{Miyakesc},
which could account for the high temperature
superconductivity in PuCoGa$_5$.
\emph{Acknowledgment}: We would like to thank G. Lander and A. Hiess for
numerous discussions of this problem, and for providing us the raw
data of his scattering experiments which are plotted in this work. We
thank M. Dzero for an early collaboration in the initial stage of this
work.  The work of M. Pezzoli and G. Kotliar was supported by BES
DOE-grant BES-DOE Grant DE-FG02-99ER45761.  K. Haule acknowledges the
support of ACS Petroleoum Research Fund 48802 and Alfred P. Sloan
foundation.

\bibliography{paper}

\newpage

\onecolumngrid
\setcounter{equation}{0}
\renewcommand{\theequation}{S\arabic{equation}}

\begin{center}
\begin{Large}
\textbf{Supplementary Material}
\end{Large}
\end{center}

\section{Form factor in second quantization}
  
The neutron magnetic form factor can be evaluated in terms of current $\vJ(\vr)$ by
the following formula
\begin{eqnarray}
\label{Fm}
\vF_M(\vq)=\frac{i m_e}{e  \hbar } \frac{1}{q^2} \vq\times \langle \vJ(\vq)\rangle \,,
\end{eqnarray}
where $m_e$, $e$ are respectively the electron mass and the electron charge ($e < 0$), $\vq$
is the scattering wave vector and
\begin{equation}
\vJ(\vq) = \int d\vr e^{-i\vq \cdot \vr} (\vJ_{\textrm{orb}}(\vr)+\vJ_{\textrm{spin}}(\vr)) \,.
\end{equation}
The orbital and spin currents  in terms of the field operators $\Psi_s(\vr)$ are given by
\begin{eqnarray}
\vJ_{\textrm{orb}}(\vr) &=& \frac{e\hbar}{2 i m_e}\sum_{s}
\left[\Psi^\dagger_s(\vr) (\vnabla\Psi_{s}(\vr)) - (\vnabla\Psi^\dagger_s(\vr)) \Psi_{s}(\vr)\right]\\
\vJ_{\textrm{spin}}(\vr) &=& \frac{e\hbar}{2 m_e}\sum_{ss'}
\left[\Psi^\dagger_s(\vr) (\vnabla\Psi_{s'}(\vr)) + (\vnabla\Psi^\dagger_s(\vr)) \Psi_{s'}(\vr)\right] \times \vec{\sigma}_{ss'} \, .
\end{eqnarray}
Note that the Fourier transform of the spin current $\vJ_{\textrm{spin}}(\vq)$ greatly simplifies: in fact
integration by parts leads to cancellation of the derivatives of the field
operator and we obtain $\vJ_{\textrm{spin}}(\vq)= -e\hbar i/(2m_e)\sum_{ss'} \vec{\sigma}_{ss'}\rho_{ss'}(\vq) \times \vq$.\\
Inserting the expression for the currents into Eq.~(\ref{Fm}), we get
\begin{equation}
\vF_M(\vq) = \frac{1}{q^2}\sum_{ss'}\int d\vr e^{-i\vq\vr}
\Psi^\dagger_s(\vr)\;\vq\times\left[\frac{1}{2}\vec{\sigma}_{ss'}\times\vq + \delta_{ss'}\vnabla\right]
\Psi_{s'}(\vr)
\end{equation}
Now we express the field operator in terms of a complete set of one
particle wave functions, such as Kohn-Sham orbitals
\begin{equation}
\Psi_s(\vr) = \sum_{i\in bnd, \vk\in 1BZ} \psi_{k i}(\vr,s) \hat{c}_{\vk i} \,,
\end{equation}
to get
\begin{eqnarray}
\vF_M(\vq) = \frac{1}{q^2}\sum_{\vk,ij,ss'} n^{DMFT}_{\vk,ij}
\int_{cell} d\vr
e^{-i\vq\vr}
\psi^*_{\vk i}(\vr,s)\;\vq\times\left[\frac{1}{2}\vec{\sigma}_{ss'}\times\vq + \delta_{ss'}\vnabla\right]\, .
\psi_{\vk j}(\vr,s')
\end{eqnarray}
Here $n^{DMFT}_{\vk,ij}$ is the DMFT density matrix expressed in the
Kohn-Sham base, $\vk$ runs over the first Brillouin zone only, and $i,j$
run over Kohn-Sham bands.  Because $\vq$ is reciprocal vector, the
integration over space is performed only over one unit cell (denoted
by $cell$).
When the spin-orbit coupling is large, bands do not have the spin
index, because there is mixing between both spins species, hence we
need to double the number of bands.\\
In the LAPW basis, there are two contributions to the above equation:
within muffin-tin and in the interstitial region.

\subsection{Within Muffin-Tin}

 Inside the Muffin-Tin the KS orbitals are expressed in terms of coefficients
$A_{i,\vK s}^\vk$ and LAPW basis functions
$\chi_{\vk+\vK}(\vr,s)$ as
\begin{equation}
\psi_{\vk i}(\vr,s)=\sum_{\vK }A_{i,\vK s}^\vk \chi_{\vk+\vK}(\vr,s) =
\sum_{\vK s} A_{i,\vK s}^\vk\; a^{\vk\kappa t}_{L\vK}\; u_l^{\kappa t}(r_t)\;Y_{L}(\hat{r}_t)\chi_s\, ,
\end{equation}
where $\vK$ are reciprocal lattice vectors, $t$ marks the atom type, 
$u^{0 t}_l(r)$, $u^{1 t}_l(r)$, $u^{2 t}_l(r)$ are the radial solutions of the Dirac equation,
its energy derivative, and optional local orbitals; $Y_L(\hat{r}_t) $ are
spherical harmonics with $L$ labeling the angular quantum numbers $l,m$ .\\
For shorter notation, we define a new type of density matrix in the muffin-tin subspace
\begin{equation}
n^{DMFT}_{\vk,t, L s \kappa, L's'\kappa'} = \sum_{ij}
\left(\sum_\vK A^{\vk\;*}_{i,\vK s}\; a^{\vk\kappa t\; *}_{L \vK}\right)
\; n^{DMFT}_{\vk, ij}\;
\left(\sum_{\vK'} A^{\vk}_{j,\vK' s'}\; a^{\vk\kappa' t}_{L' \vK'}\right)
\end{equation}
and we express the form factor inside the muffin-tin  in terms of this density matrix
\begin{eqnarray}
\vF_M(\vq) = \frac{1}{q^2}\sum_{\vk,t,L s \kappa, L' s'\kappa'}
n^{DMFT}_{\vk, t, L s \kappa, L's'\kappa'}
\langle u_l^\kappa\; Y_L\; \chi_s|
e^{-i\vq\vr}
\left[
\frac{1}{2} \;\vq\times (\vec{\sigma}_{ss'}\times\vq) +
\delta_{ss'}
\vq\times \vnabla\right]
| u_{l'}^{\kappa'}\; Y_{L'}\; \chi_{s'}\rangle_t
\end{eqnarray}
Here integration runs over muffin-tin sphere $t$.\\
From the above expression it is not obvious that the $q\rightarrow 0$ is
well behaved. However, we can add any constant to exponent
$e^{i\vq \cdot \vr}$ in the orbital part of the expression, because it
vanishes due to symmetry. It is therefore possible to write an
alternative expression
\begin{eqnarray}
\vF_M^{MT}(\vq) = \frac{1}{q^2}\sum_{\vk,t,L s \kappa, L' s'\kappa'}
n^{DMFT}_{\vk, t, L s \kappa, L's'\kappa'}
\langle u_l^\kappa\; Y_L\; \chi_s|
\left[
\frac{1}{2}e^{-i\vq\vr}
\;\vq\times (\vec{\sigma}_{ss'}\times\vq) +
\delta_{ss'}
(e^{-i\vq\vr}-1)
\vq\times \vnabla\right]
| u_{l'}^{\kappa'}\; Y_{L'}\; \chi_{s'}\rangle_t
\end{eqnarray}
which clearly is well behaved in the $\vq=0$ limit.

\subsection{Interstitial Region}

In the interstitial region the KS solution is
\begin{equation}
\psi_{\vk i}(\vr ,s) = \sum_{\vK }
A^{\vk}_{i,\vK s} \frac{1}{\sqrt{V_{cell}}} C_{\vK} e^{i(\vk+\vK)\vr} \chi_s
\end{equation}
We again define a corresponding density matrix
\begin{equation}
n^{DMFT}_{\vk,\vK s,\vK' s'} =\sum_{ij}
A^{\vk\; *}_{i,\vK s} C_{\vK}^* n^{DMFT}_{\vk,ij}
A^{\vk}_{j,\vK' s'} C_{\vK'}
\end{equation}
and express the form factor by
\begin{equation}
\begin{split}
\vF_M^I(\vq)& = \frac{1}{q^2}\sum_{\vk,\vK s,\vK' s'}
n^{DMFT}_{\vk, \vK s, \vK' s'} \\
\times & \left[
\frac{1}{2}
\;\vq\times (\vec{\sigma}_{ss'}\times\vq) +
i\,\delta_{ss'}\,
\vq\times(\vk+\frac{1}{2}\vK + \frac{1}{2}\vK')\right]
\left[
\delta_{\vK'-\vK-\vq}-
\sum_t \frac{4\pi R_t^2}{V_{cell}}\frac{j_1(|\vK'-\vK-\vq|R_t)}{|\vK'-\vK-\vq|}
\right]
\end{split}
\end{equation}
The last term comes from the difference of the integral over the
entire unit cell and inside all muffin-tin spheres.\\

\section{Dipole approximation}

In the dipole approximation, we approximate the Fourier exponent with
\begin{equation}
e^{-i\vq \,\vr}\approx j_0(q r) -i (\vq\cdot\vr) (j_o(q r)+j_2(q r))
\end{equation}
and obtain
\begin{eqnarray}
\vF_M(\vq) = \frac{1}{q^2}\sum_{\vk,ij,ss'} n^{DMFT}_{\vk,ij}
\int_{cell} d\vr
\left[j_0(q r) -i (\vq\cdot\vr) (j_o(q r)+j_2(q r))\right]
\psi^*_{\vk i}(\vr,s)\;\vq\times\left[\frac{1}{2}\vec{\sigma}_{ss'}\times\vq + \delta_{ss'}\vnabla\right]
\psi_{\vk j}(\vr,s').
\end{eqnarray}
Due to parity selection rules, only  the following two terms are nonzero
\begin{equation}
\begin{split}
\label{II2}
\vF_M(\vq) & = \frac{1}{q^2}\sum_{\vk,ij,ss'} n^{DMFT}_{\vk,ij} \\
\times & \int_{cell} \; d\vr
\psi^*_{\vk i}(\vr,s)
\left[
j_0(q r)\frac{1}{2}\vq\times(\vec{\sigma}_{ss'}\times\vq)
 -i \delta_{ss'}\,[j_o(q r)+j_2(q r)](\vq\cdot\vr)(\vq\times  \vnabla)
\right]
\psi_{\vk j}(\vr,s').
\end{split}
\end{equation}
Inside the expression Eq.~(\ref{II2}) for form factor, we can use
$$-i(\vq\cdot\vr)(\vq\times\vec{\nabla})= \frac{1}{2} \frac{m_e}{\hbar} \frac{\textrm{d}}{\textrm{dt}}\big[(\vq \cdot \vr)(\vq \times \vr) \big]
- \frac{1}{2}i (\vq \cdot \vr)(\vq \times \vnabla) + \frac{1}{2}i(\vq \cdot \vnabla)( \vq \times \vr) \,. $$
The following term vanishes
\begin{equation}
\sum_{\vk, ij} n^{DMFT}_{\vk, ij}
\langle\psi_{i\vk}|\frac{1}{2} \frac{m_e}{\hbar} \frac{\textrm{d}}{\textrm{dt}}\big[(\vq \cdot \vr)(\vq \times \vr) \big] |\psi_{j\vk}\rangle=
\left \langle\frac{1}{2} \frac{m_e}{\hbar} \frac{\textrm{d}}{\textrm{dt}}\big[(\vq \cdot \vr)(\vq \times \vr) \big] \right \rangle= 0 \,.
\end{equation}
Hence we can use 
\begin{equation}
 - \frac{1}{2}i (\vq \cdot \vr)(\vq \times \vnabla) + \frac{1}{2}i(\vq \cdot \vnabla)( \vq \times \vr) =
\frac{i}{2} \vq \times [ \vq \times( \vr \times \vnabla)]
\end{equation}
and $\vr \times \vnabla = \frac{i}{\hbar} \vl $ to write
\begin{equation}
 i(\vq\cdot\vr)(\vq\times\vec{\nabla}) = \frac{1}{2 \hbar} \vq \times ( \vl \times \vq)\,;
\end{equation}
Finally we find the expression for the form factor in the dipole approximation
\begin{equation}
\begin{split}
\vF_M(\vq) & = \frac{1}{q^2}\sum_{\vk,ij,ss'} n^{DMFT}_{\vk,ij} \\
\times & \int_{cell} \; d\vr
\psi^*_{\vk i}(\vr,s)
\left[
\frac{1}{2}\vq\times(\vec{\sigma}_{ss'}\times\vq)j_0(q r)
 +\delta_{ss'} \frac{1}{2 \hbar}
[ (\vq\times\vec{l}) \times \vq]
\,[j_o(q r)+j_2(q r)] 
\right]
\psi_{\vk j}(\vr,s').
\end{split}
\end{equation}
Inside the muffin-tin the Kohn Sham orbitals are expressed in terms of
a basis in which $\vec{l}$ and $\vec{s}$ are diagonal. 
Therefore the form factor in the dipole approximation, inside the muffin-tin, reads
\begin{eqnarray}
\vF_M(\vq) = 
\sum_{\vk,t,sL,\kappa,\kappa'}
n^{DMFT}_{\vk,t,L s \kappa, L s\kappa'}
\frac{\vq\times(\vec{e}_z\times\vq)}{q^2}
 \left[s_z
\langle u_l^\kappa|j_0(q r)|u_{l}^{\kappa'}\rangle
+ \frac{1}{2} l_z\,
\langle u_l^\kappa|j_o(q r)+j_2(q r)|u_{l}^{\kappa'}\rangle \,
\right].
\end{eqnarray}
where we took the magnetic field in $z$ direction.

\section{Numerical evaluation}

For numerical evaluation we split the form factor expression into
the dipole part and the correction to the dipole approximation $\Delta
F_M$. The correction is 
\begin{eqnarray}
\Delta \vF_M(\vq) = \frac{1}{q^2}\sum_{\vk,ij,ss'} n^{DMFT}_{\vk,ij}
\langle \psi_{\vk i,s}|
\left[e^{-i\vq\,\vr}-j_0(q r)\right]\frac{1}{2}\vq\times(\vec{\sigma}_{ss'}\times\vq)
 + \delta_{ss'}\,
 \left[e^{-i\vq\,\vr}+i(\vq\cdot\vr)\left(j_o(q r)+j_2(q r)\right)\right](\vq\times  \vnabla)
|\psi_{\vk j,s'}\rangle\nonumber.
\end{eqnarray}
Inside the muffin-tin sphere, this expression takes the form
\begin{eqnarray}
\Delta \vF_M(\vq) =
\frac{1}{q^2}
\sum_{\vk,t,s,L\kappa,L'\kappa'} n^{DMFT}_{\vk,Ls\kappa t,L's\kappa' t}
\left\{
\vq\times(\vec{e_z}\times\vq)
\frac{1}{2}\sigma^z_{ss}
\langle u_l Y_L|
\left[e^{-i\vq\,\vr}-j_0(q r)\right]
|u_{l'}Y_{L'}\rangle
\right.\\
\left.
+\langle u_l Y_L|
 \left[e^{-i\vq\,\vr}-1+i(\vq\cdot\vr)\left(j_o(q r)+j_2(q r)\right)\right](\vq\times  \vnabla)
|u_{l'}Y_{L'}\rangle
\right\}
\nonumber.
\end{eqnarray}
We compute in advance the following quantities
\begin{eqnarray}
a_{LL'}(r) &=& \langle Y_L|e^{-i\vq\,\vr}-1|Y_{L'}\rangle\\
b_{LL'}(r) &=& \langle Y_L|-i\vq\cdot\vr|Y_{L'}\rangle\\
\vec{c}_{LL'}(r)&=& \langle Y_L|\vq\times\vec{\nabla}|Y_{L'}\rangle\\
\vec{d}_{LL'}(r)&=& \langle Y_L|\vq\times \vec{e}_{\vr}|Y_{L'}\rangle
\end{eqnarray}
and obtain the following one dimensional integral over the radial
distance in the muffin-tin sphere
\begin{eqnarray}
\Delta \vF_M(\vq) =
\frac{1}{q^2}
\sum_{\vk,t,s,L\kappa,L'\kappa'} n^{DMFT}_{\vk,Ls\kappa t,L's\kappa' t}
\left\{
\vq\times(\vec{e_z}\times\vq)
\frac{1}{2}\sigma^z_{ss}
\int dr r^2 u_l^\kappa(r) u_{l'}^{\kappa'}(r) [a_{LL'}(r)+\delta_{LL'}\left(1-j_0(qr)\right)]
\right. \\
+\int dr r^2 u_l^\kappa(r) \frac{d u_{l'}^{\kappa'}(r)}{dr}
\sum_{L''}\left\{a_{LL''}(r) - b_{LL''}(r)\left[j_0(qr)+j_2(qr)\right]\right\}\vec{d}_{L'' L'}(r)
   \nonumber \\
\left.
+\int dr r^2 u_l^\kappa(r) u_{l'}^{\kappa'}(r)
\sum_{L''}\left\{a_{LL''}(r) - b_{LL''}(r)\left[j_0(qr)+j_2(qr)\right]\right\}\vec{c}_{L'' L'}(r)
\right\}
\nonumber.
\end{eqnarray}
To check numerical accuracy, one could check the accuracy of the
following sum
\begin{eqnarray}
\sum_{L''} b_{LL''}\vec{c}_{L'' L'} = \langle Y_L|-i(\vq\cdot\vr)(\vq\times\vec{\nabla})|Y_{L'}\rangle=\nonumber\\
-i\langle Y_L|(\vq\times\vr)(\vq\cdot\vec{\nabla})-i\vq\times(\vq\times\vec{l})|Y_{L'}\rangle=\nonumber\\
-i\langle Y_L|(\vq\times\vr)(\vq\cdot\vec{\nabla})|Y_{L'}\rangle\nonumber
-\vq\times(\vq\times\vec{e_z})l_z \delta_{LL'}\nonumber\,;
\end{eqnarray}
since for $L=L'$ the first term is zero,  we have
\begin{eqnarray}
\sum_{L''} b_{LL''}\vec{c}_{L'' L} =-\vq\times(\vq\times\vec{e_z})l_z
\end{eqnarray}
To derive the above equation, it is useful to know the following property of
the spherical harmonics
\begin{eqnarray}
\frac{d}{d\theta}Y_{l,m}(\theta,\phi) =
\frac{1}{2}\sqrt{(l-m)(l+m+1)}e^{-i\phi}Y_{l,m+1}(\theta,\phi) -
\frac{1}{2}\sqrt{(l+m)(l-m+1)}e^{i\phi}Y_{l,m-1}(\theta,\phi)  
\end{eqnarray}
which follows from
\begin{eqnarray}
\sqrt{1-x^2}\frac{d}{dx}P_{l,m}(x) = -\frac{1}{2} P_{l,m+1}(x)+\frac{1}{2}(l+m)(l-m+1)P_{l,m-1}(x) \,.
\end{eqnarray}

\section{Magnetic susceptibility for $\textrm{PuCoGa}_5$}

From the calculation of the magnetic form factor $F_M(q)$ we can extract the 
magnetic susceptibility as $\chi = \mu/B$, where
$\mu = - 2 \mu_B F_M(0)$ is the magnetic moment of the Pu atom and $B$ is the  applied magnetic field.
The results are summarized in table \ref{Tabsusc}.
\begin{center}
\begin{table}[h!]
\caption{\label{Tabsusc} Magnetic susceptibility for PuCoGa$_5$ obtained with the NCA impurity solver.}
 \begin{tabular}{c|c}
\hline
  T (Kelvin) & $\chi$ ($\mu_B/\textrm{Tesla}$) \\
\hline
\hline
  12.5     & $12 \times 10^{-4}$ \\
   25      & $13 \times 10^{-4}$ \\
   50      & $15 \times 10^{-4}$ \\
\hline
 \end{tabular}
\end{table}
\end{center}
The computed magnetic susceptibility compares well  with the values obtained 
by the neutron experiment [ A. Hiess \emph{et. al.}, Phys. Rev. Lett. {\bf 100}, 076403 (2008)].
DFT+DMFT obtains a larger susceptibility that the measured one, which is comprehensible
since it is well known that the Non Crossing Approximation (NCA) impurity solver
underestimates the Kondo temperature.\\
To solve the impurity problem in  presence of an applied magnetic field
we gave special attention to the off-diagonal terms in the impurity hybridization
strength $\Delta$, which give a significant contribution to the form factor.

\end{document}